\documentclass[preprint,amsmath,amssymb]{revtex4-1}

\usepackage{dcolumn}
\usepackage{bm}
\usepackage{url}
\usepackage{graphicx}
\usepackage{color}
\usepackage{amssymb}
\usepackage{latexsym}
\usepackage{multirow}
\usepackage{enumerate}
\usepackage{twoopt}
\usepackage{array}
\usepackage{alltt}
\usepackage[center]{subfigure}

\definecolor{orange}{rgb}{0.8,0.2,0.2}

\newcommand{\up}{\vspace{-0.2cm}}
\newcommand{\down}{\vspace{0.2cm}}

\newcommand{\Down}{\vspace{0.4cm}}

\newcommand{\tbf}[1]{\textbf{#1}}

\newcommand{\reals}{\mathbb{R}}
\newcommand{\mbf}[1]{\mathbf{#1}}

\newcommand{\mathdef}{\stackrel{\triangle}{=}} 

\newcommand{\hamiltonian}{\mathcal{H}}

\newcommand{\mtcspace}{\mathcal{S}}
\newcommand{\setmtcspace}{\mathcal{Z}}

\newcommand{\netspace}{\mathcal{N}}

\newcommand{\loglambda}{\log_{10}(\lambda)}

\newcommand{\pos}[1][k]{\mbf{r_{#1}}}
\newcommand{\mom}[1][k]{\mbf{p_{#1}}}

\newcommandtwoopt{\sample}[2][][]{s_{#1}^{#2}}
\newcommand{\netsample}[1][]{n_{#1}}
\newcommand{\distance}[1][ij]{d_{#1}}
\newcommand{\netlink}[1][ij]{L_{#1}}

\newcommand{\kmeans}{\emph{k}-Means}
\newcommand{\LS}{\emph{LS}}
\newcommand{\TS}{\emph{TS}}

\begin{document}

\title{Datasets as Interacting Particle Systems: a Framework for Clustering}

\author{Giuliano Armano}
\email{armano@diee.unica.it}
\affiliation{DIEE - Dept. of Electrical and Electronic Engineering\\ University of Cagliari, P.zza D'armi 09123
Cagliari, Italy}

\author{Marco Alberto Javarone}
\email{marco.javarone@diee.unica.it}
\affiliation{DIEE - Dept. of Electrical and Electronic Engineering\\ University of Cagliari, P.zza D'armi 09123
Cagliari, Italy}

\begin{abstract}
In this paper we propose a framework inspired by interacting particle physics and devised to perform clustering on multidimensional datasets.
To this end, any given dataset is modeled as an interacting particle system, under the assumption that each element of the dataset corresponds to a different particle and that particle interactions are rendered  through gaussian potentials.
Moreover, the way particle interactions are evaluated depends on a parameter that controls the shape of the underlying gaussian model. In principle, different clusters of proximal particles can be identified, according to the value adopted for the parameter.
This degree of freedom in gaussian potentials has been introduced with the goal of allowing multiresolution analysis.
In particular, upon the adoption of a standard community detection algorithm, multiresolution analysis is put into practice by repeatedly running the algorithm on a set of adjacency matrices, each dependent on a specific value of the parameter that controls the shape of gaussian potentials. 
As a result, different partitioning schemas are obtained on the given dataset, so that the information thereof can be better highlighted, with the goal of identifying the most appropriate number of clusters.
Solutions achieved in synthetic datasets allowed to identify a repetitive pattern, which appear to be useful in the task of identifying optimal solutions while analysing other synthetic and real datasets. 

\end{abstract}

\maketitle

\section{Introduction}
Complex networks are used in different domains to model specific structures or behaviors. Relevant examples are the Web, biological neural networks, and social networks Ref.~\cite{albert02,sporns04,guimer03}.
Community detection is one of the most important processes in complex network analysis, aimed at identifying groups of highly mutually interconnected nodes, called communities Ref.~\cite{newman04}, in a relational space. 

From a complex network perspective, a community is identified after modeling the given data as a graph. 
For instance, a social network inherently contains communities of people linked by some (typically binary) relations --e.g., based on friendship, sports, hobbies, movies, books, or religion.
On the other hand, from a machine learning perspective, a community can be thought of as a cluster of elements.
In this case, elements of the given domain are usually described by a set of features, or properties, which permit to assign each instance a point in a multidimensional space.
The concept of similarity is prominent here, as clusters are typically identified by focusing on common properties (e.g., age, employment, health records). 

Although complex networks are apparently suited to deal with relations rather than properties, we deem they could also be used for partitioning multidimensional datasets --characterizing themselves as an alternative to classical clustering algorithms.
To reach this goal, we used a metaphor taken from particle systems physics.
Indeed, resorting to theoretical physics for setting up complex networks algorithms and/or for studying their properties has often been helpful for getting new insights and for devising effective methods.
Just to cite few, Bianconi and Barabasi Ref.~\cite{bianconi01} defined physical models for dynamical networks, comparing Bose-Einstein Condensation to winner-takes-all policies. Barabasi Ref.~\cite{albert02} showed how tools of statistical mechanics can be useful in the study of complex networks. Kriukov et al. Ref.~\cite{krioukov01} developed a geometric framework to study the structure and function of complex networks, interpreting edges as non-interacting fermions whose energies are hyperbolic distances between nodes.
Gudkov et al. Ref.~\cite{Nussinov08} devised and implemented a method for detecting communities and hierarchical substructures in complex networks. The method represents nodes as point masses in an $N-1$ dimensional space and uses a linear model to account for mutual interactions.

In this paper we propose a framework for clustering multidimensional datasets, in which data samples are represented as interacting particles. The framework has been tested on synthetic and real datasets.
The remainder of the paper is organized as follows: Section~\ref{sec:dataset-interacting-particle} gives a brief introduction to the physics of interacting particle systems and introduces the model. Section~\ref{sec:framework} describes the proposed framework.
Section~\ref{sec:kmeans} focuses on $\kmeans$, a centroid-based clustering algorithm used to further validate the results obtained with the proposed framework.
Section~\ref{sec:exp-results} reports experimental results.
Conclusions (Section~\ref{sec:conclusions}) end the paper.

\section{Viewing Datasets as Interacting Particle Systems} \label{sec:dataset-interacting-particle}

\subsection{Physics of Interacting Particle Systems}

Let us consider a system with $n$ mutually interacting identical particles, each with mass $m$. Solids, liquids and gases are examples of particle systems.
An overall description of the system can be given by its Hamiltonian $\hamiltonian$. If the system is isolated, we know that $\hamiltonian \equiv E = const$ (with $E$ denoting its internal energy).
A viable way for understanding the behavior of any such system is to analyze intermolecular interactions. To this end, classical dynamics can be used, under the assumption that molecules are chemically inert and that the forces among molecules depend only on the reciprocal distance. Under these assumptions we can write $\hamiltonian$ as:
\begin{equation}
\hamiltonian = \sum_{i=1}^{N}{\frac{\mom[i]^{2}}{2m}} + \sum_{i=1}^{N-1}\sum_{j=i+1}^{N}{u(\pos[i], \pos[j])}
\end{equation}

\down

\noindent where $\mom[i]$ and $\pos[i]$ are the momentum and the radius vector of the $i$-th particle, respectively (see Ref.~\cite{Landau01}).

\subsection{Modeling Datasets}
The aim of the research activity described in this paper was to devise and implement a framework for clustering multidimensional datasets without a priori knowledge about them.
To deal with this problem, still open in the machine learning community (see, for example, Ref.~\cite{Jain10,Tibshirani01,Hansen98}), we took a cue from particle physics.
To this end, we decided to model datasets as interacting particle systems, using a simplified form of the Hamiltonian, in which the kinetic energy equals to zero and the potential energy contains an attractive component.
As attractions depend only on the distance among elements, it becomes viable to groups atoms according to their proximity relations. Of course, in the proposed model, an atom/molecule corresponds to an element of the dataset and vice versa.

To put the model into practice, one must define the potential among elements. In particular, we defined a gaussian family of functions that computes the attractive potential between two elements.
In doing so, a dataset is codified by a complex network, whose nodes denote the elements of the dataset (i.e., particles) and whose links denote their attractive potentials. Overall, proximity values give rise to an adjacency matrix.
Notably, different adjacency matrices can be generated for the same dataset (i.e., more than one complex network), depending on the function used to evaluate the proximity among particles.

\section{Clustering Framework} \label{sec:framework}

The proposed framework, called IPMC (Interacting Particles Model for Clustering), has been devised to perform clustering on multidimensional datasets by means of a multiresolution complex network analysis. The underlying conjecture is that complex network analysis can became an appropriate tool also in the field of machine learning.

\subsection{Computing interactions among elements}

Let us briefly recall that a metric space is identified by a set $\setmtcspace$ of dimensions, together with a distance function 
$\distance[]: \setmtcspace \times \setmtcspace \rightarrow\ \reals$, like Euclidean, Manhattan and Chebyshev distances.
In IPMC, the underlying assumption is that a sample $\sample$ can be described by $N$ features $f_{1},f_{2}, ..., f_{N}$, encoded as real numbers. In other words, the sample can be represented as a vector in an $N$-dimensional metric space, say $\mtcspace$. 
Our goal is to give rise to a fully connected weighted network starting from $\mtcspace$ while taking into account  the distance function that holds in that space.
Conversely, the complex network space will be denoted as $\netspace$ hereinafter, with the underlying assumption that for each sample $\sample[i] \in \mtcspace$ a corresponding element $\netsample[i] \in \netspace$ exists and vice versa. 
This assumption makes easier to evaluate the proximity value $\netlink[ij]$ between two elements $\netsample[i], \netsample[j] \in \netspace$, according to the distance $\distance[ij]$ between the corresponding elements $\sample[i], \sample[j] \in \mtcspace$.

Without loss of generality, let us assume that each feature in $\mtcspace$ is normalized in $[0,1]$ and that a function $\psi : \reals \rightarrow \reals$ exists for computing the interactions among elements in~$\netspace$, starting from the value of the distance function in $\mtcspace$. In symbols:
\begin{equation} \label{eq:metric-to-network}
\netlink[](\netsample[i],\netsample[j]) = \netlink[ij] \mathdef \psi(\distance[ij]) = \psi(\distance[](\sample[i], \sample[j]))
\end{equation}
Evaluating interactions for all pairs of samples in $\netspace$ (i.e., evaluating their weighted links) gives rise to a fully connected complex network.
Moreover, recalling that $\mtcspace$ is normalized in $[0,1]$, we expect $\netlink[ij] \approx 0$ when $\distance[ij] \approx \sqrt{N}$, $N$ being the number of features of the space $\mtcspace$. The value $\sqrt{N}$ comes from the following inequality, which holds for any pair of samples $\sample[i], \sample[j] \in \mtcspace$ (represented by their vector representation in terms of the given set of features $\pos[i],\pos[j]$):

\begin{equation}
\distance[ij] = \sqrt{ \sum_{k=1}^N  (\pos[i][k] - \pos[j][k])^2} \le  \sqrt{N} \end{equation}

\noindent where $\pos[i][k]$ denotes the $k$-th component of $\pos[i]$.

\subsection{The Adopted Community Detection Algorithm}

Community detection is the process of finding communities in a graph (the process is also called ``graph partitioning'').
As stated in Ref.~\cite{fortunato10}, identifying communities is feasible only when the graph is sparse, i.e., when 
$m \approx n$, where $m$ denotes the number of links and $n$ the number of nodes.
From a computational perspective, this is not a simple task and many algorithms have been proposed, according to three main categories: divisive, agglomerative, and optimization algorithms.
In our work, we used the Louvain method Ref.~\cite{blondel08}, an optimization algorithm based on an objective function devised to estimate the quality of partitions.
In particular, at each iteration, the Louvain Method tries to maximize the so-called \emph{weighted-modularity}, defined as:
\begin{equation} \label{eq:weighted-modularity}
Q = \frac{1}{2m} \cdot \sum_{i,j}\left [A_{ij} - \frac{k_{i}k_{j}}{2m} \right] \cdot \delta(\sample[i],\sample[j])
\end{equation}
\noindent where $A_{ij}$ is the generic element of the adjacency matrix, $k$ is the degree of a node, $m$ is the total ``weight'' of the network, and $\delta(s_{i},s_{j})$ is the Kronecker Delta, used to assert whether a pair of samples belongs to the same community or not.

\subsection{Multiresolution Analysis}

To perform a multiresolution analysis on the network space, a parametric family $\Psi(\lambda): \reals \rightarrow \reals$ of functions is required, where $\lambda$ is a parameter that controls the shape of each concrete $\psi$ function. After setting a value for $\lambda$, the corresponding $\psi$ can be used to convert the distance computed for each pair of samples in the given dataset into a proximity value. 
In particular, the following parametric family of Gaussian functions has been experimented:

\begin{equation} \label{eq:gaussian-family}
\Psi(\lambda;x) = e^{-\lambda x^2}
\end{equation}

Hence, the weight of the link between two nodes $\netsample[i], \netsample[j] \in \netspace$, i.e., $\netlink[ij]$, can be evaluated according to Equation~\eqref{eq:gaussian-family}:

\begin{equation} \label{eq:metric-to-network-1}
\netlink[ij] \mathdef \psi(\lambda;\distance[ij]) = e^{-\lambda \distance[ij]^2}
\end{equation}

\noindent where the $\lambda$ parameter is used as a constant decay of the link. In doing so, the Hamiltonian of our model is defined as:
\begin{equation} \label{eq:hamiltonian_ipmc}
\hamiltonian =
\sum_{i=1}^{N-1}\sum_{j=i+1}^{N}{u(\pos[i], \pos[j])} =
\sum_{i=1}^{N-1}\sum_{j=i+1}^{N} e^{-\lambda \distance[ij]^2}
\end{equation}

Following the definition of $\Psi(\lambda;x)$ as $e^{-\lambda x^2}$, multiresolution analysis takes place varying the value of the $\lambda$ parameter. The specific strategy adopted for varying $\lambda$ is described in the experimental section (Section~\ref{sec:exp-results}). It is worth noting in advance that an exponential function with negative constant decay ensures that distant points in an Euclidean space are loosely coupled in the network space and vice versa. 
\down

\section{Centroid-based clustering} \label{sec:kmeans}

Experimental results obtained with the proposed method have been compared with those obtained by running a classical clustering algorithm.
As centroid-based clustering is one of the most acknowledged clustering strategies, the $\kmeans$ algorithm (e.g., Ref.~\cite{alsabti98}), which belongs to this family, has been selected as comparative tool. For the sake of completeness, let us briefly summarize it:

\Down

\begin{minipage}{14cm}
\footnotesize
\begin{enumerate}
  \setlength{\itemsep}{1pt}
  \setlength{\parskip}{0pt}
  \setlength{\parsep}{0pt}
\item Place k centroids in the given metric space;
\item Assign each sample to the closest centroid,
   thus identifying tentative clusters;
\item Compute the Center of Mass (CM) of each cluster;
\item IF CMs and centroids (nearly) coincide THEN STOP;
\item Let CMs become the new centroids;
\item REPEAT from STEP 2.
\end{enumerate}
\normalsize
\end{minipage}

\Down

The evaluation function of $\kmeans$, called \emph{distortion} and usually denoted as $J$, is computed according to the formula:
\begin{equation} \label{eq:distortion}
J = \sum_{j=1}^{k} \sum_{i=1}^{n_j} \left | \sample[i][(j)] - c_{j}\right |^{2}
\end{equation}
\noindent where $n_j$ is the number of samples that belong to the $j$-th cluster, $\sample[i][(j)]$ is the $i$-th sample belonging to $j$-th cluster, and $c_{j}$ its centroid.
Note that different outputs of the algorithm can be compared in terms of distortion only after fixing $k$, i.e., the number of clusters.
In fact, comparisons performed over different values of $k$ are not feasible, as the more $k$ increases the lower the distortion is. For this reason, the use of $\kmeans$ entails a main issue: how to identify the ``right'' number $k$ of centroids (see Ref.~\cite{eick04}). 

\section{Experimental Results} \label{sec:exp-results}

Experimental procedure has been developed in three main phases: i) a preliminary phase, aimed at performing IPMC on few and relatively simple synthetic datasets, ii) a test phase, aimed at performing IPMC on other, more complex, datasets, and iii) a comparative phase, aimed at comparing the behavior of IPMC and $\kmeans$.
\subsection{Preliminary phase}
A first group of $4$ synthetic datasets, called \LS~(i.e., Learning Set) hereinafter, has been generated. Their main characteristics are summarized in Table~\ref{tab:feature-datasets1}. IPMC has been run on these datasets.

\begin{table*}[!ht]
\renewcommand\arraystretch{1.0} 
\centering
\begin{tabular}{|c| c c c | c c|}
\hline
\multicolumn{1}{| m{1.3cm} | }{\begin{center}\emph{Group}\end{center}} &
\multicolumn{1}{m{1.2cm}}{\begin{center}$Dim$\end{center}}  &
\multicolumn{1}{m{1.2cm}}{\begin{center}$N_{s}$\end{center}}  &
\multicolumn{1}{m{1.2cm} |}{\begin{center}$N_{c}$\end{center}} &
\multicolumn{1}{m{1.2cm}}{\begin{center}$\mu_{r}$\end{center}} &
\multicolumn{1}{m{1.2cm} |}{\begin{center} $\sigma_{r}$ \end{center}}\\ \hline \hline
\tbf{\LS} 
& 2D &1897 &5 & $0.4$ &$0.3$ \\
& 3D &1683 & 3 & $0.09$ &$0.04$ \\
& 3D & 1500 & 10 & $0.42$ &$0.22$ \\
& 4D &1680 & 6 & $0.62$ &$0.45$\\
\hline
\end{tabular}
\caption {\label{tab:feature-datasets1} Characteristics of datasets used in the preliminary experimental phase. \emph{Dim}, \emph{$N_{s}$}, and $N_{c}$ denote the dimension of datasets, the number of samples, and the intrinsic number of clusters. Moreover, $\mu_{r}$ and $\sigma_{r}$ denote the average radium and the variance of samples.}
\end{table*}
Figure~\ref{fig:example1} shows the 3D datasets, with 3 and 10 clusters, respectively, with their optimal solutions achived by IPMC.
\begin{figure*}
\label{fig:example1}
\centering
\includegraphics[width=2.1in]{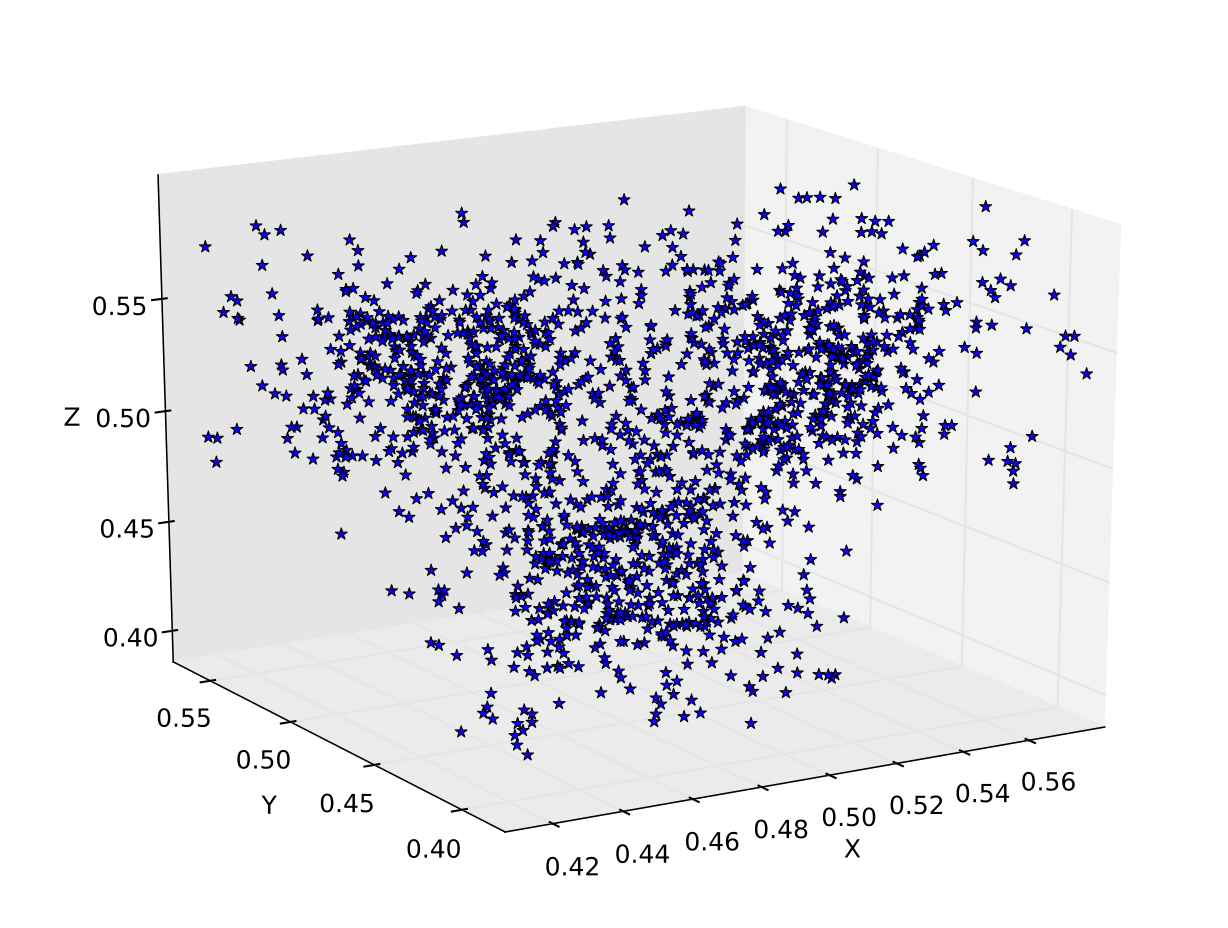}%
\includegraphics[width=2.1in]{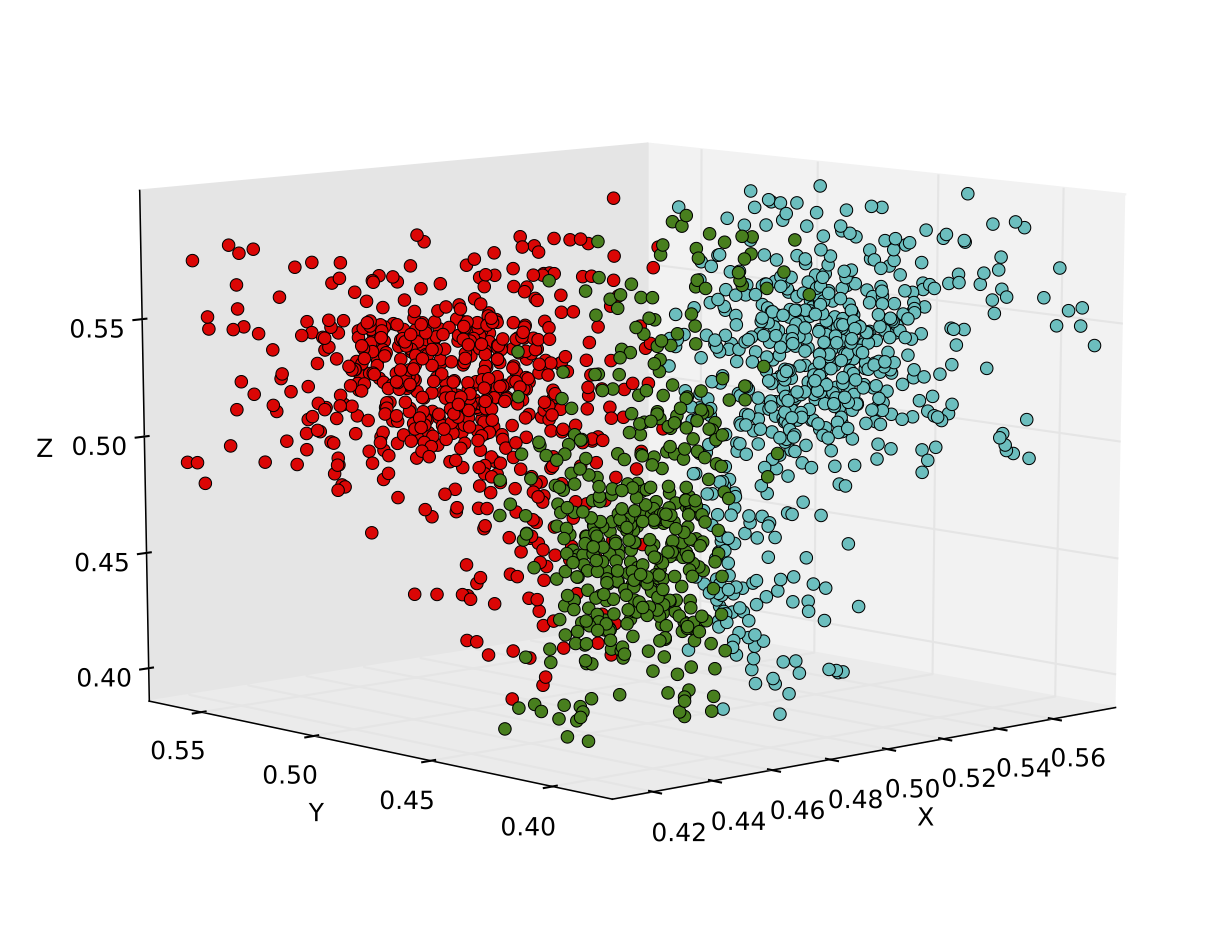}\\
\includegraphics[width=2.1in]{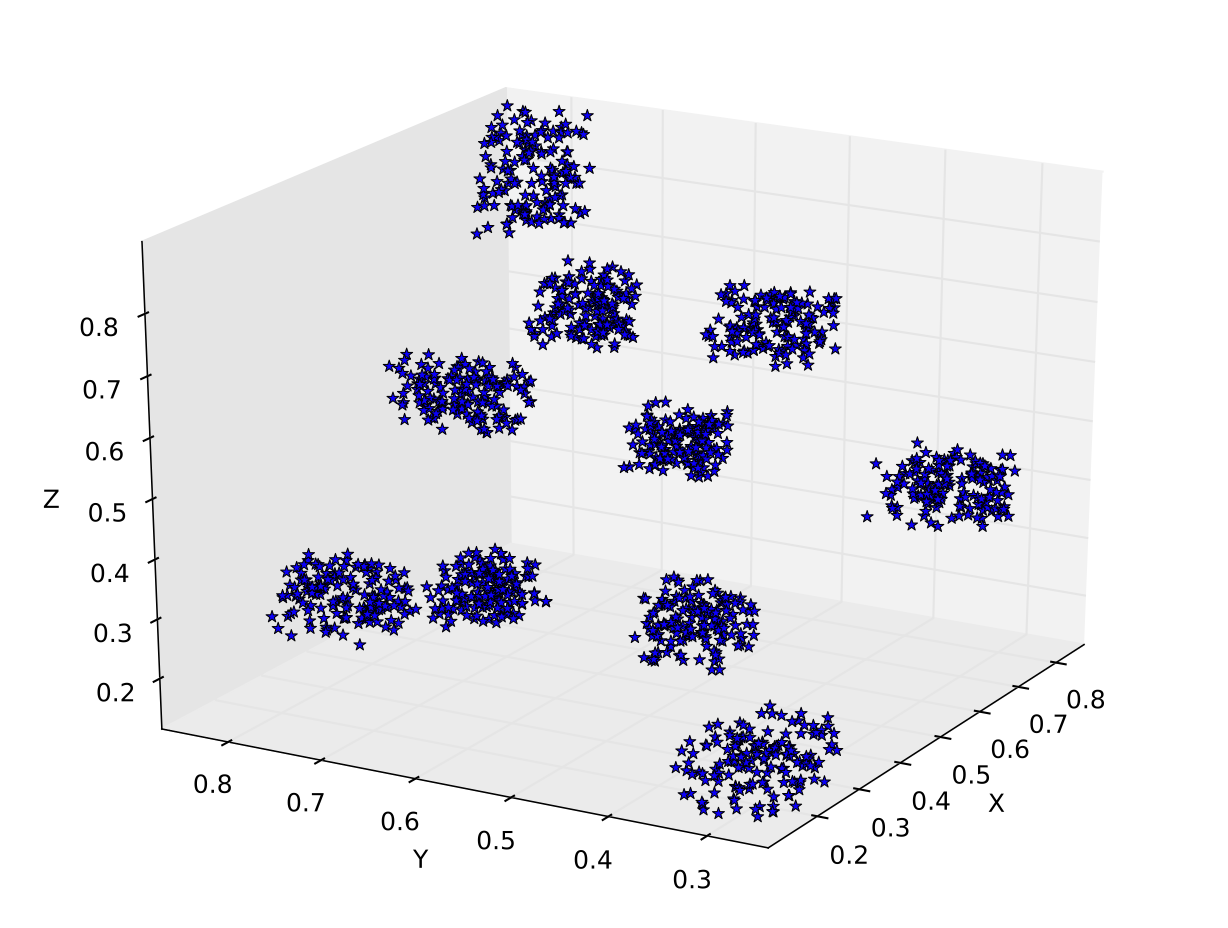}%
\includegraphics[width=2.1in]{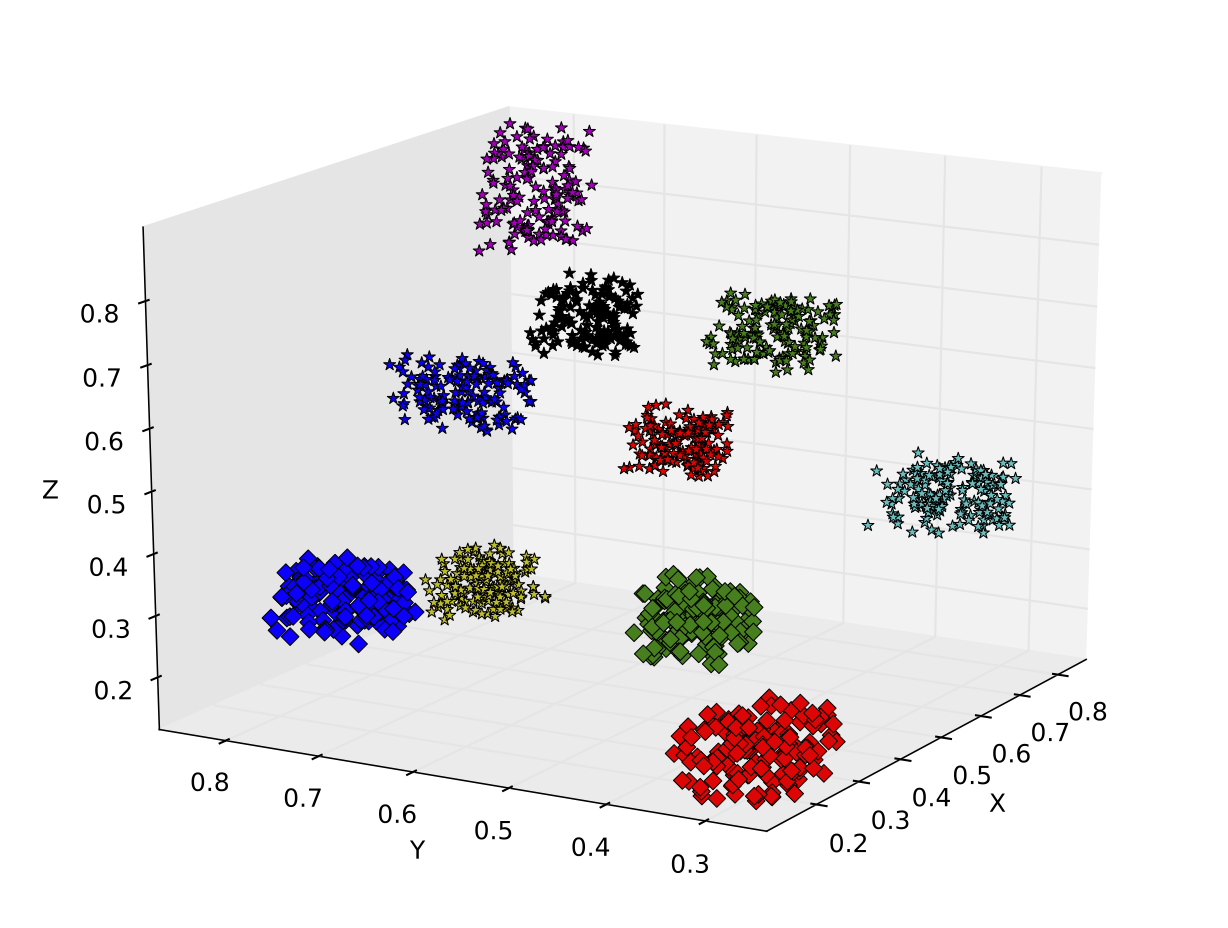}
\caption{\label{fig:example1} Second and third datasets of \LS, together with the solutions achieved by IPMC using $\loglambda = 3$ (each cluster has been colored with a different color).}
\end{figure*}
Multiresolution analysis has been performed varying the value of $\lambda$ according to Equation~\eqref{eq:gaussian-family}. A logarithm scaling has been used for $\lambda$, as we experimentally found that small changes had a negligible impact on the corresponding algorithm for community detection. In particular, for each dataset, we calculated the adjacency matrix for all values of $\lambda$ such that $\loglambda = 0, 1, 2, 3, 4$. It is worth pointing out that the maximum value of $\loglambda$ is expected to depend on the cardinality of the dataset in hand --the greater the cardinality, the greater the value of $\loglambda$. However, for most datasets, a value of $\loglambda = 4$, i.e., $\lambda = 10,000$, appears to be large enough to include all relevant information by means of multiresolution analysis. 
Table~\ref{tab:result-datasets1} shows the results of multiresolution analysis of the preliminary phase.
\begin{table*}[!ht]
\renewcommand\arraystretch{1.0} 
\centering
\begin{tabular}{|c| c | c c c c c |}
\hline
\multicolumn{1}{| m{1.1cm} | }{\begin{center}\emph{Group}\end{center}} &
\multicolumn{1}{m{1.1cm} |}{\begin{center}$N_{c}$\end{center}} &
\multicolumn{5}{ m{5.0cm}|}{\begin{center}~~Number of Clusters  \end{center}} \\ \hline \hline
\tbf{\LS} 
&5 & ~~2~~ & ~~3~~ & ~\tbf{5}~ & ~\tbf{5}~ & \tbf{5}\\
& 3&\tbf{3}  & \tbf{3}  & \tbf{3} & \tbf{3} & 103\\
& 10 &2 & 3 & \tbf{10} & \tbf{10} & 151\\
& 6 &2 & 4 & \tbf{6} & \tbf{6} & 37\\
\hline
\hline
&& 0 & 1 & 2 & 3 & 4\\
&&\multicolumn{5}{ m{5.0cm}|}{\up \begin{center}~~$\loglambda$\end{center}} \\
\hline
\end{tabular}
\caption {\label{tab:result-datasets1} Results of multiresolution analysis on the selected datasets. The table reports the number of communities, calculated for $\loglambda = 0, 1, 2, 3, 4$. Optimal values are reported in bold.}
\end{table*}
A non trivial problem, here, is how to identify the optimal $\lambda$.
Fortunately, IPMC always allows to identify optimal or suboptimal solutions\footnotemark~while performing multiresolution analysis on \LS~datasets.
\footnotetext{As pointed out by Arenas et al. Ref.~\cite{arenas01}, it may not appropriate to speak of correct vs. incorrect solutions for multiresolution analysis. In a context of community detection we deem more appropriate to speak of optimal or suboptimal solutions (i.e., see also Ref.~\cite{zhifang01} for more information on this issue).}
In particular, we observed the following pattern to occur: the optimal solution is robust with respect to major changes of $\lambda$ (let us recall that this parameter is in the logarithm scale). Indeed, it is achieved by more then one value of $\loglambda$. Our hypothesis was that this pattern can be considered as a decision rule aimed at identifying the optimal $\lambda$.
\subsection{Test phase}
Considering the analysis made on \LS~dataset, we generated a second group of datasets, characterized by an increasing complexity with respect to \LS. This second group of datasets is denoted as \TS, (i.e., Testing Set) hereinafter. Also on these new datasets, we performed IPMC, with the aim of verifying the validity of the identified pattern. Still with the intention of assessing the pattern, we performed experiments using \emph{Iris}, the most famous real dataset available at Ref.~\cite{iris}. 
Iris is a well known multivariate dataset, containing 50 samples described by 4 attributes, from each of 3 species of Iris (Iris setosa, Iris virginica and Iris versicolor).
Table~\ref{tab:feature-datasets2} summarizes the main characteristics of \TS and \emph{Iris}.\\
\begin{table*}[!ht]
\renewcommand\arraystretch{1.0} 
\centering
\begin{tabular}{|c| c c c c |c c|}
\hline
\multicolumn{1}{| m{1.3cm} | }{\begin{center}\emph{Group}\end{center}} &
\multicolumn{1}{m{1.2cm}}{\begin{center}$Dim$\end{center}}  &
\multicolumn{1}{m{1.2cm}}{\begin{center}$S/R$\end{center}}  &
\multicolumn{1}{m{1.2cm}}{\begin{center}$N_{s}$\end{center}}  &
\multicolumn{1}{m{1.2cm}}{\begin{center}$N_{c}$\end{center}} &
\multicolumn{1}{| m{1.2cm}}{\begin{center}$\mu_{r}$\end{center}} &
\multicolumn{1}{m{1.2cm} |}{\begin{center} $\sigma_{r}$ \end{center}}\\ \hline \hline
\tbf{\TS}
& 3D & S & 350 & 5 & $0.35$ &$0.19$\\
& 3D & S & 2000 & 20 & $0.44$ &$0.2$ \\
& 3D & S &5000 & 30 & $0.51$ &$0.24$\\
& 4D   & S &535 & 4 & $0.64$ &$0.46$ \\
& 8D   & S &1680 &6 & $0.86$ &$0.62$ \\
& 12D & S &930 & 8 & $1.22$ &$0.88$ \\
\hline
\tbf{Iris}
& 4D  & R  &150 & 3 & $0.49$ &$0.26$\\
\hline
\end{tabular}
\caption {\label{tab:feature-datasets2} Characteristics of datasets listed out according to the group they belong to. \emph{Dim}, \emph{S/R}, \emph{$N_{s}$}, and $N_{c}$ denote the dimension of datasets, whether the dataset is synthetic (S) or real (R), the number of samples, and the intrinsic number of clusters. Moreover, $\mu_{r}$ and $\sigma_{r}$ denote the average radium and the variance of samples.}
\end{table*}
The corresponding results, obtained with IPMC, are shown in Table~\ref{tab:result-datasets2}.
\begin{table*}[!ht]
\renewcommand\arraystretch{1.0} 
\centering
\begin{tabular}{|c| c | c | c c c c c |}
\hline
\multicolumn{1}{| m{1.1cm} | }{\begin{center}\emph{Group}\end{center}} &
\multicolumn{1}{m{1.1cm} |}{\begin{center}$N_{c}$\end{center}} &
\multicolumn{1}{m{1.4cm} |}{\begin{center}Pattern\end{center}} &
\multicolumn{5}{ m{5.0cm}|}{\begin{center}~~Number of Clusters \end{center}} \\ \hline \hline
\tbf{\TS}
& 5 &\checkmark  &3 & \tbf{5} & \tbf{5} & 8 & 84\\
& 20 &\checkmark  &3 & 4 & 16 & \tbf{20} & \tbf{21}\\
& 30 &\checkmark  &4 & 5 & 21 & \tbf{30} & \tbf{30}\\
& 4  &\checkmark &2 & \tbf{4}  & \tbf{4}  & 105 & 181\\
& 6  &\checkmark &2 & 4 & \tbf{6}  & \tbf{6} & 1186\\
& 8 &\checkmark  &3 & 5 & \tbf{8}  & \tbf{8} & 875\\
\hline
\tbf{Iris}
& 3  & \checkmark & \tbf{3} & \tbf{3}  & 10 & 82 & 147\\
\hline
&&  & 0 & 1 & 2 & 3 & 4\\
&&  & \multicolumn{5}{ m{5.0cm}|}{\up \begin{center}~~$\loglambda$\end{center}} \\
\hline
\end{tabular}
\caption {\label{tab:result-datasets2} Results of multiresolution analysis on the selected datasets, listed out according to the group they belong to. The table reports the number of communities, calculated for $\loglambda = 0, 1, 2, 3, 4$. Optimal values are reported in bold. Occurring pattern, observed on synthetic datasets (and reported in the table for the sake of completeness) allows to easily compute the expected optimal number of communities also for \emph{Iris}.}%
\end{table*}
Looking at these results, we still observe the pattern identified in the preliminary phase.
Furthermore, we observed that a correlation often exists between the cardinality of the dataset in hand and the order of magnitude of its optimal $\lambda$ (typically, the former and the latter have the same order of magnitude). It is also interesting to note that in some datasets of \LS~(i.e., 2nd, 3rd and 4th) and of \TS~(i.e., 4th, 5th and 6th) the optimal $\lambda$ precedes a rapid increase in the number of communities.
As a final note, we found no significant correlation between the optimal $\lambda$ and the weighted-modularity parameter, notwithstanding the fact that this parameter is typically important to assess the performance of the adopted community detection algorithm.
\subsection{Comparison phase: IPMC vs $\kmeans$}  
\begin{figure*}
\centering
\includegraphics[width=4in]{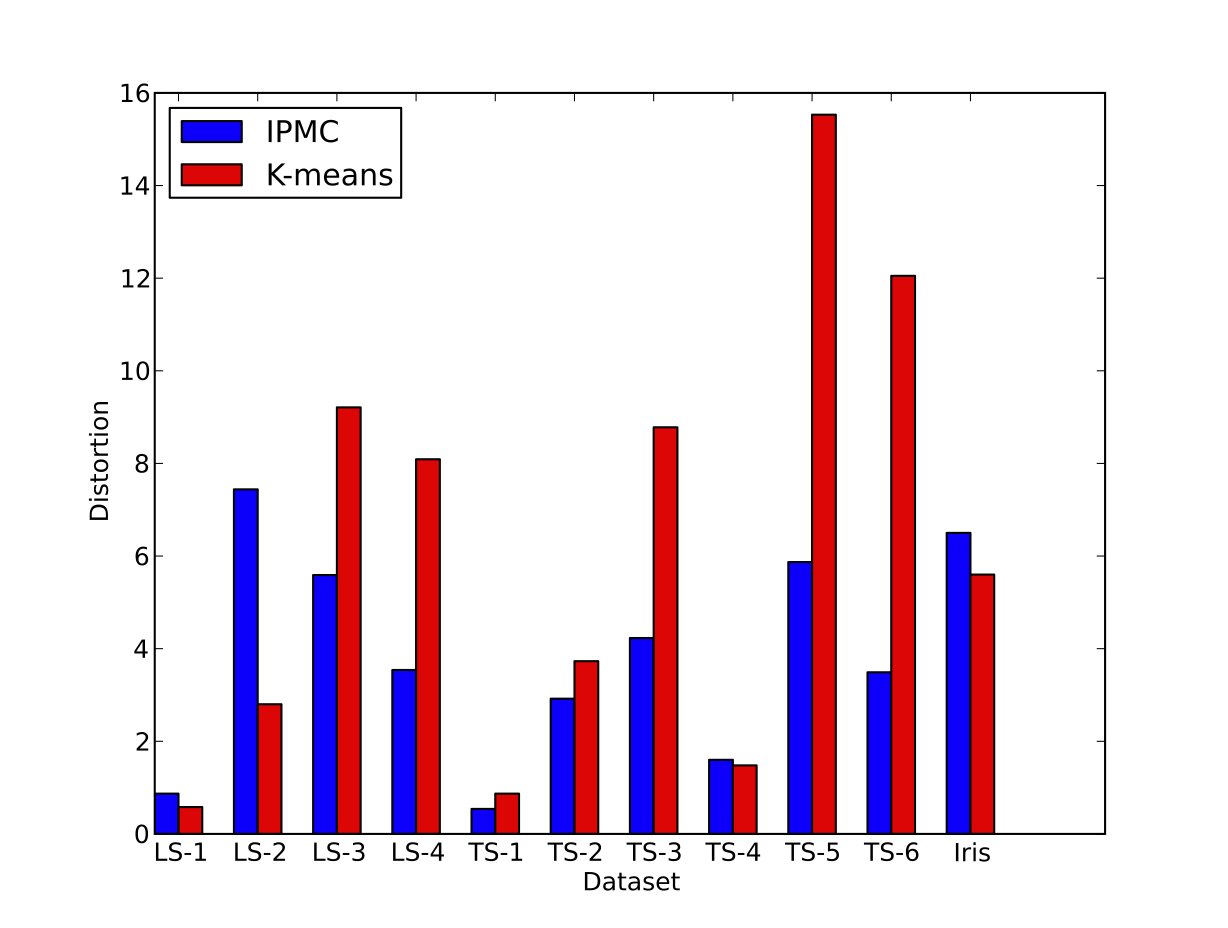}
\caption{\label{fig:distortion} Comparison, in terms of distortion, between solutions achieved by IPMC, blue bars, and $\kmeans$, red bars (the less the better).}
\end{figure*}

For the sake of comparison, we decided to run the \kmeans~algorithm (using the Euclidean metric) on the selected datasets, with the goal of getting new insights on the results of the partitioning procedure defined in IPMC.
It is worth pointing out that the algorithm has been run using the optimal values of $k$ identified by means of the multiresolution analysis.
Figure~\ref{fig:distortion} reports comparative results and clearly shows that, in the $64$ percent of the cases, IPMC computes a better result than $\kmeans$. This result highlights the validity of the proposed framework, also considering that IPMC computes partitions without any a priori knowledge about the datasets, as the optimal (or suboptimal) number of clusters is typically found by applying the previously described pattern. Although $\kmeans$ is faster than IPMC, it is important to stress that its results, at each attempt, depend tightly on the initial position of the $k$ centroids. Hence, in absence of a strategy for identifying the initial disposal of centroids, $\kmeans$ should be (and it is in fact) run several times, the solution with the smaller distortion being selected as optimal.
\section{Conclusions} \label{sec:conclusions}
In this paper, a framework for clustering multidimensional datasets has been described, able to find the most appropriate number of clusters also in absence of a priori knowledge.
Viewing a dataset as an interacting particle system, we have shown that community detection can be effectively used also for data clustering tasks and that results are comparable with those achieved by a classical clustering algorithm.   
The proposed framework makes use of transformations between metric spaces and enforces multiresolution analysis.
A comparative assessment with a well-known clustering algorithm (i.e., \kmeans) has also been performed, showing that IPMC often computes better results.
As for future work, we are planning to test IPMC with other relevant datasets, focusing on two main issues: verify further on its validity and compare it with other relevant clustering algorithms (in particular, with algorithms able to estimate the optimal or suboptimal number of clusters in advance). Furthermore, we are planning to study to which extent one can rely on the decision pattern described in the paper, assessing its statistical significance on a large number of datasets.

\section*{Acknowledgments}
Many thanks to Alessandro Chessa and to Vincenzo De Leo (both from Linkalab). The former for his wealth of ideas about complex networks and the latter for the support given to install and run their Complex Network Library. 

\bibliography{bibliography}

\end{document}